# Noninvasive Realistic Stimulation/Recording of Freely Swimming Weakly Electric Fish: Movement Detection and Discharge Entropy to Infer Fish Behavior

Caroline G. Forlim[1,2], Reynaldo D. Pinto[2]

[1] Lab. Fenômenos Não-Lineares, Instituto de Física, Universidade de São Paulo – Cx. Postal 66318, 05315-970, São Paulo, SP, Brazil

[2] Lab. Neurodinâmica/Neurobiofísica, Instituto de Física de São Carlos/USP - Cx. Postal 369, 13560-970, São Carlos, SP, Brazil

contact e-mail: carolineforlim@usp.br, reynaldo@ifsc.usp.br

**short title:** Realistic Stimulation of Weakly Electric Pulse Fish.

**Acknowledgements:**
This work was supported by the Brazilian agencies: FAPESP, CAPES, and CNPq.

**List of abbreviations:**

ADC - Analog to Digital Converter
DAC - Digital to Analog Converter
EOD – Electric Organ Discharge
IPI – Inter Pulse Interval
LIA - Local Induced Amplitude
JAR - Jamming Avoidance Response
NR - Novelty Response
PSTH - Post Stimulus Time Histograms

# Abstract

Weakly electric fish are unique models in Neuroscience allowing experimentalists to access, with non invasive techniques, a central nervous system generated spatio-temporal electric pattern of pulses with roles in at least two complex and not yet completely understood abilities: electrocommunication and electrolocation. We developed an apparatus to allow realistic stimulation and simultaneous recording of electric pulses in freely moving *Gymnotus carapo* for very long periods (several days). Voltage time series from a three-dimensional array of sensitive dipoles that detects electric field in several positions underwater were digitized and home made real-time software allowed reliable recording of pulse timestamps, independently of the fish's position, and also to infer fish movement. A stimulus fish was mimicked by a dipole electrode that reproduced the voltage time series of real conspecific pulses, but according to timestamp sequences previously recorded that could be chosen by the experimenter. Two independent variables were used to analyze fish behavior: the entropy of the recorded timestamp sequences and the movement of the fish inferred from pulse amplitude variability at each detection dipole. All fish presented very long transient exploratory behavior (about 8 hours) when exposed to a new environment in the absence of stimuli. After the transient there were several intervals (from 5 min to 2 hours), in which entropy vanished and no movement was observed, that could be associated with behavioral sleeping. Our experiments also revealed that fish are able to discriminate between real and random stimuli distributions by changing the timing probability of the next discharge. Moreover, most fish presented behavioral sleep periods when the artificial fish timestamp sequence was random, but no fish showed any behavioral sleep period when the artificial fish fired according to a real fish timestamp series.

# Introduction

Weakly electric fish offer an unique opportunity to study and interact with living intact nervous systems with non-invasive methods since their electric organ discharges (EODs) are not only complex signals that belong to the processing of environmental sensory information but they are also easy-to-detect (Lissmann, 1958; Bullock, 1999; Fortune et al., 2006, Engelmann et al., 2008).

The production of electric signal in pulse-type weakly electric fish has two basic components, the waveform and the timing of the electric pulses (Lissmann, 1958). The pulse waveforms are species-specific and the inter EOD intervals, or inter pulse intervals (IPIs), can vary from 10 to hundreds of milliseconds (Caputi, 2004). Each EOD generates electric fields that propagates into the water, interact with the surroundings, and reach a myriad of electroreceptors disposed to detect changes in the electric field that cause small alterations in self or conspecific transcutaneous currents (Bullock, 1999; Caputi, 2004).

Fish are able to localize and measure distance of nearby objects with different impedance from that of the surrounding water by analyzing the alterations suffered in the self-generated electric field, in a process called active electrolocation (von der Emde et al., 1998, 2008, 2010; Caputi et al., 2002; von der Emde and Fetz, 2007; Engelmann, 2008). They also perform electrocommunication, in which one fish's EODs interfere with the other fish's receptors. During this task, each fish emits and receives self and conspecific generated EODs (Caputi and Budelli, 1995; von der Emde, 1999; Doiron et al., 2003). Neuroethology of electrolocation and electrocommunication remain as interesting and active fields of research (Engelmann et al., 2010; Ho et al., 2010; von der Emde et al., 2010; Cuddy et al., 2011).

In many pulse-type species, such as Gymnotus carapo studied here, the EOD present among individuals a stereotyped pulse shape with no sexual dimorphism (Caputi et al., 2002). In such species the electrocommunication must be realized by modulating the IPIs. Moreover, fish can also vary their IPIs depending on the environmental context: when a new stimulus (light, sound, electrical field, vibration) is presented a transient acceleration of the EOD rate, called novelty response (NR) (Post and von der Emde, 1999; Aguilera and Caputi 2002) is triggered. There was also been reported a mechanism of jamming avoidance response (JAR) in which closer fish slightly adjust their EOD timing to avoid discharging at the same time and interfere with each others' electrolocation (Heiligenberg et al., 1978; Capurro e Malta, 2004).

The EODs generate a complex spatio-temporal pattern that is experimentally difficult to deal with in freely moving fish (Westby, 1975). This is the main reason, in spite of the sensitivity of the fish IPI patterns to stimuli, environment, and manipulation, many previous experiments to characterize the electrosensory system were conducted restricting the fish movements. While deeply interfering with the fish normal behavior and consequently with the informational content of IPIs precluded studying the social interaction among fish, such experiments were invaluable to the knowledge of the mechanisms involved in electrolocation (Caputi et al., 2002; Bullock et al. 2005; Hitschfeld et al. 2009). Studies of electrical interactions between fish were mostly possible in species that presented, at least during breeding season, an accentuated sexual dimorphism in EOD shape (Wong and Hopkins, 2007; Perrone et al., 2009) or in cases where specimens with very different sizes were used (Westby, 1975; Westby, 1979; Arnegard and Carlson, 2005).

In this context we developed an apparatus that allowed non invasive, realistic and controlled stimulation with the simultaneous recording of electric pulses in freely moving Gymnotus carapo for very long periods of time (several days). We concentrate our experiments in two different situations: (1) non stimulated fish; (2) fish receives stimuli from an artificial fish that mimics the pulse shape of a conspecific. We could successfully address: how long the stress caused by exposing the fish to a new environment distorts the patterns of electrical activity, the changes in the electrical activity when exposed to real (conspecific) or random distributions of stimuli IPIs, the presence of a JAR mechanism, and the use of inferred entropy and movement of the fish to classify its behavior.

# Materials and methods

## Animals

Adult specimens, 15-25cm long, of unsexed *Gymnotus carapo* were acquired from local fishermen in the state of São Paulo, Brazil. The fish were maintained in individual 30 liter (30x35x30 cm) unplanted tanks exposed to natural illumination at room temperature and furnished with some lengths of PVC pipe (hiding places). Water conductivity in all experimental installations was in the range 100±5 μS/cm, similar to that found in the streams/lagoons where fish are usually captured. While in the lab, fish were fed twice a week on a variety of living food: small fish, worms, and *artemia salina* (brine shrimp). A total of 13 fish were used in the experiments reported, some of them were recorded several times in different protocols.

All experimental protocols and procedures followed the ethical principles suggested by the Society for Neuroscience and were approved by the Committee on Ethics in Animal Experimentation of the Federal University of São Carlos – UFSCar (Protocol number 007/2009).

## Measurement aquarium

The experimental setup, totally developed in our laboratory (Pinto et al., 2007, Forlim, 2008, Forlim et al., 2011) consists of a 64 liters glass aquarium (40 x 40 x 45 cm) (Fig. 1), enclosed in a double Faraday cage (1 mm aluminum sheets) shielding electrical noise, and two 3,0 cm thick layers of insulating foam confining a 4,0 cm air layer to attenuate sound waves. Other possible low amplitude, low frequency mechanic vibrations were attenuated by hanging the whole heavy apparatus on steel cables attached to the ceiling of the building.

## Artificial stimulus generation

A dipole electrode inside a 15 cm PVC pipe was used to generate artificial stimuli mimicking the geometry of a medium size fish. Artificial fish stimuli were generated by a Digital to Analog Converter (DAC) board (Digidata 1200B, Axon Instruments, Union City, CA), controlled by real-time software developed by our research group (Pinto et al., 2001, Nowotny et al., 2006) to mimic a real *Gymnotus carapo* EOD waveform, previously stored in a PC compatible computer (Fig. 2A), and rescaled to a fixed peak amplitude of 5V. Two stimuli timestamp sequences (30 min long) could be chosen by the experimenter: one pre-recorded from an unstimulated real fish, and one with random (flat distribution) generated inter-pulse intervals (IPI) in the same range of the real fish (between 15 and 20 ms).

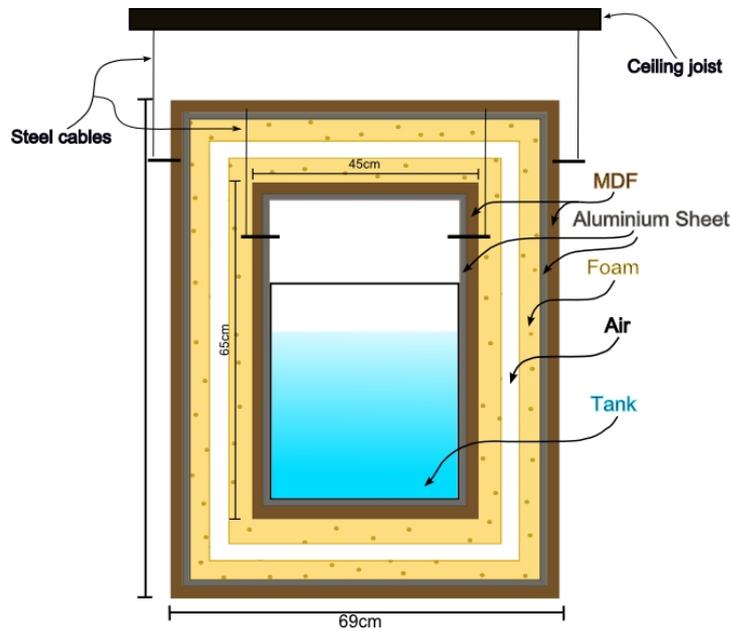

**Figure 1**. **Measurement aquarium.** The experimental setup, developed in our own laboratory, consists on a 64 liters glass aquarium (40x40x45 cm). Two boxes made of fiberboard (MDF, 20 mm thick) were used as support for the apparatus. A double Faraday cage (1 mm aluminum sheets) shields electrical noise and two 3 cm thick layers of insulating foam confining a 4,0 cm air layer attenuates sound waves. Possible low amplitude, low frequency mechanic vibrations were attenuated by hanging the whole heavy apparatus on steel cables attached to the ceiling of the building.

## Multi-electrode detection system

Spatio-temporal EOD was measured using a three-dimensional array of 8 electrodes, each one consisting of a 0.2 mm diameter stainless steel wire tightly inserted through silicone glue, between glass aquarium's walls, dipping typically 1-2 mm inside the water. The electrodes were placed in a cubic arrangement (~ 40 cm on a side, Fig. 2A). Signals from seven electrodes were differentialy amplified (gain = 100x, home-made) using the electrode at the base of the aquarium as a common reference.

Time series of all differential amplifiers, containing both EOD and artificial stimulus signals, were digitized at 50 kHz by an analog to digital converter (ADC) board (PCI MIO 16-E1, National Instruments, Texas) in an acquisition PC compatible computer.

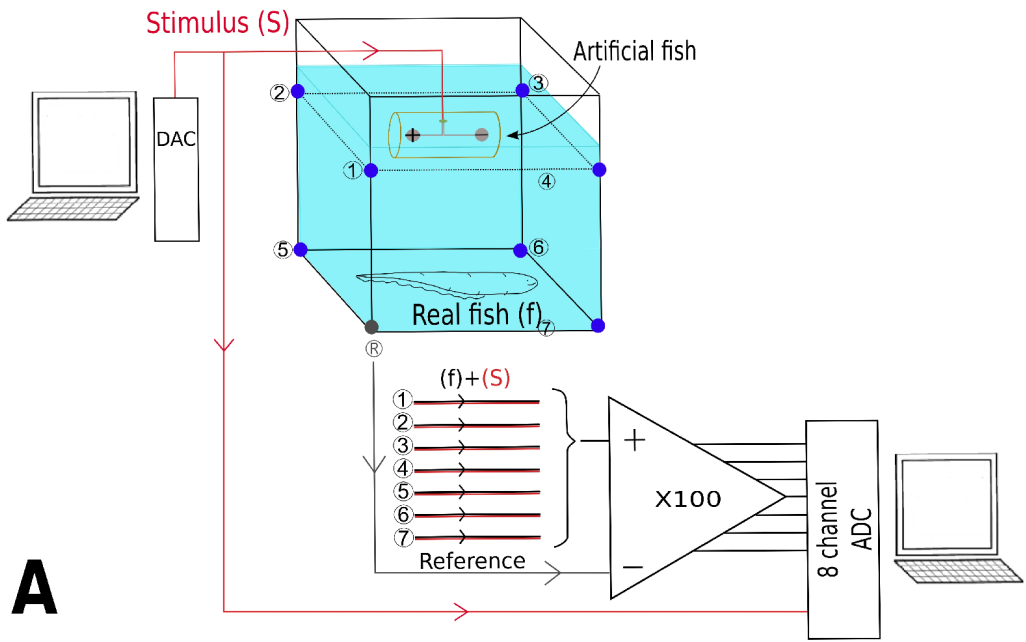

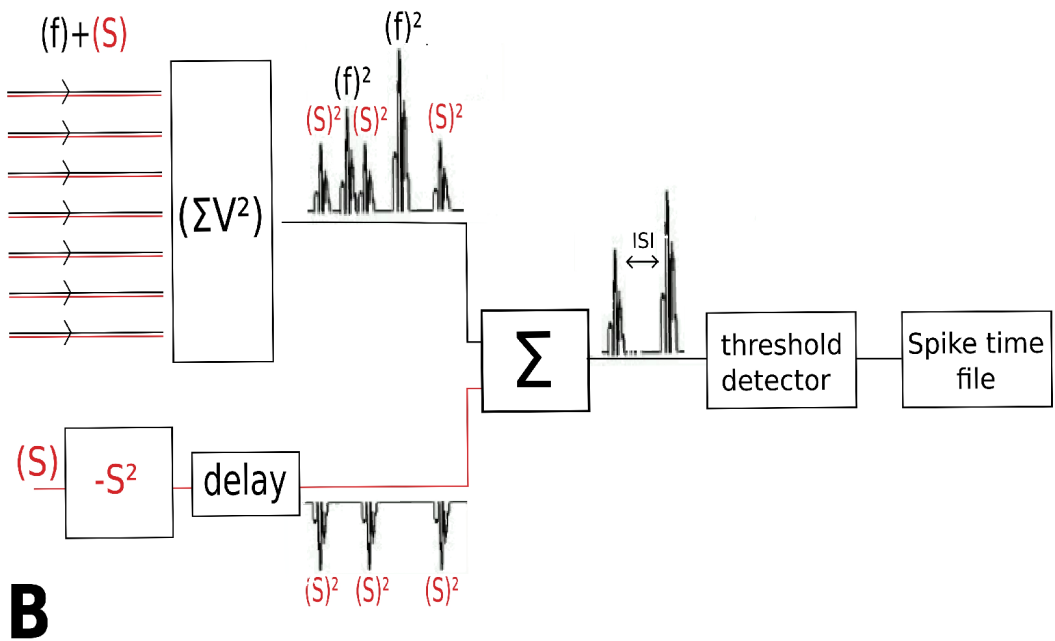

**Figure 2. Acquisition and stimulation setup. A- hardware:** EODs from real (f) and artificial stimulus (S) fish are measured using a 3D array of 8 electrodes ((1)-(7), and (R)). Signals are differentially amplified (x 100, (R) is the common reference) and digitized at 50 kHz by an ADC board. The artificial fish is a dipole electrode inside a 15 cm PVC pipe driven by a DAC output of a computer running real-time home made stimulation software that mimics a real fish pulse shape with IPIs defined by the experimenter. **B- acquisition software:** the square of the digitized artificial signal is delayed (to account for propagation in water and electronics) and subtracted from the squared and summed electrode signals to separate only real fish pulses. Real fish EODs are then detected with a simple threshold level and the timestamps and the amplitude of the pulses at each electrode are stored for posterior analyses.

## Acquisition software

*Gymnotus carapo* generates a multi-phasic EOD pulse (Castelló et al., 2009). A simple way to detect the timing of an EOD is to determine the occurrence of the major component of the pulse. However, depending on the fish orientation, the major component can be either negative or positive in a given electrode, making detection difficult. To overcome this problem our detection software squared and summed the digitized time series registered by all electrodes, compressing EOD information in a single time series. By simply comparing the compressed time series with a threshold level we were able to reliably detect the EODs (Fig. 2B), with a resolution of 20 µs.

For each EOD detected, the program recorded the timestamp of the event and the locally induced amplitude (LIA) at each electrode, defined here as the difference between the maximum and minimum values of voltage induced during the EOD.

In experiments with artificial stimulation, a delay was added to the ADC input of the artificial signal to account for the propagation time through water. Since the ADC input signals from the electrodes contained both real fish and artificial pulses together, we subtracted the square of the artificial signal from the sum signal to avoid detecting artificial pulses as real fish ones (Fig. 2B).

The detection software was written in Dasylab graphic language (Dasytech, Germany). Real fish timestamps, their LIAs, and artificial timestamps were stored in the acquisition computer for posterior analysis.

## Data analyses

Information Theory provides invaluable tools to study neural coding (Borst & Theunissen, 1999). Here we infer the variability of the timestamp sequences by calculating the entropy H(T) as a function of time, using a direct method (de Ruyter van Steveninck et al., 1998) adapted to our data using programs developed in our laboratory.

We analyse the timestamp sequences in T=40 s windows, but the same qualitative results were found using 20 and 60 seconds windows. The first window spans from 0 to 40 s, the second one from 20 to 60 s, and so on, maintaining a 20 s time distance between the start of successive windows (Fig. 3). In a particular window we calculate the probability of observing different EOD patterns and obtain the entropy from this probability distribution.

To compute the entropy H(T) the time axis was discretized into $b$ sec wide bins. The presence of an EOD in each bin was represented by 1 and the absence by 0. Thus we encode a sequence of EODs as a binary string (Fig. 3). The entropy $H$ depends on our choice of $b$: $H = H_b$. If $b$ is too small, most of the bits of our strings will be zero and very few patterns will be found. On the other hand, if $b$ is too big, most bits will be set to one, leading again to very few patterns. In both cases the low values of entropy will not satisfactory represent the variability of the data. Hence there must be an intermediate optimal value of $b$ able to express more correctly the variability of the data. In agreement with the principle of maximization of entropy (Shannon, 1948), choosing $b$ such that $H_b$ is a maximum will also produce the least structured set of patterns consistent with the data. Intuitively, suitable values of $b$ should provide strings with similar amounts of zeros and ones. In our case, the fish average IPI is about 20-40 ms, hence $b$ should be of the order of 10-20 ms. To ensure finding the "best" value of $b$ we compute the entropy for $b$ between 3 and 40 ms, in steps of 1 ms.

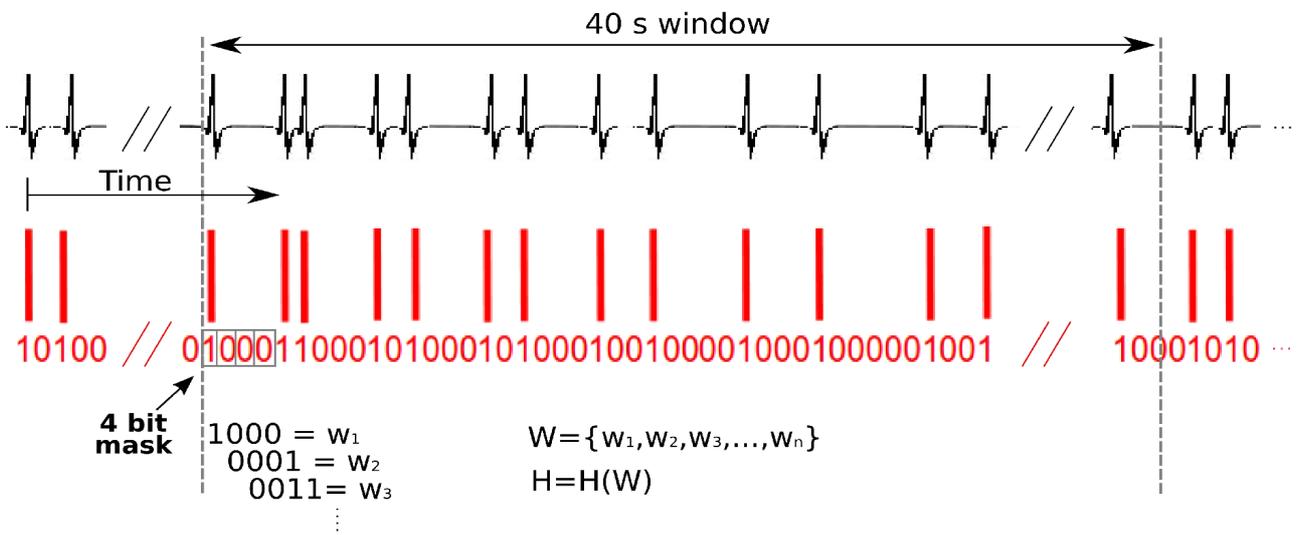

**Figure 3. Binarization and set of patterns for entropy calculations.** Time axis is discretized into small bins. The presence (absence) of an EOD in each bin is represented by 1 (0). Timestamp sequences are analyzed in bit strings obtained in 40 s windows. Each pattern is a 4 bit *word*. Starting from the beginning of the bit string a 4 bit mask is moved one bit to the right to extract a new word until the end of the window, forming the set of words $W = \{w_1, w_2, w_3,... , w_N\}$. From the set of probabilities of each word $w_i$ we compute the entropy $H(W)$.

To proceed we have to decide on the length *L* of patterns, whose probability we are going to compute. In our analysis each pattern will be represented by a *word* composed by *L*=4 bits. For a given *b* we start at the very beginning of the corresponding string and using a 4-bit mask we copy the first 4 bits to the word $w_1$. Then we move the mask one bit to the right and copy the bits to the word $w_2$ (Fig. 3). This process is repeated until the mask reaches the end of the window. $W = \{w_1, w_2, w_3,..., w_N\}$ is the set of all words (patterns) from which we compute the probabilities *P*(*w*) of finding each one of the $2^4$ = 16 different words. The entropy of *W* for that particular value of *b* is given by:

$$H_b(W) = -\sum_w P(w) \log_2 P(w)$$

To ensure a good statistical estimation of *P*(*w*) one should use stationary data and the largest window length possible. However, the whole timestamp sequences are not stationary (some of them are several days long), and we want to explore this feature by analyzing changes in the entropy along the sequence and relate them to the fish behavior. T=40 s was long enough to provide an average number of words N ~ (40 s / 15 ms) ~ 2500 words, giving a good statistics of *P*(*w*) for all possible 16 words and still capturing the fish behavior in a sub-minute scale.

Besides entropy, the electrical behavior was also characterized using IPI histograms and standard deviations of the LIAs, all calculated according to their classical definitions and for the same sequence of 40 s windows used in the entropy computation for direct comparison. To study the behavior of the fish when stimulated by different distributions of IPIs we built post stimulus time histograms (PSTHs), where we plotted the experimental probability of a fish EOD occurs immediately after a stimulus EOD (set as time reference).

The experimental setup allowed us to simultaneously address electrical and locomotor behavior of the fish. Since LIA at each electrode is strongly sensitive to the fish's position, we used the sum of standard deviations of these amplitudes to estimate the fish movement in the 40 s windows. The electrode that presented the maximum pulse amplitude was also used to infer the fish's position and orientation at each EOD.

Analysis programs were developed in C++ language and Matlab (The Mathworks Inc., Natick, MA) and ran in free open source Ubuntu Linux 10.04 (Canonical Ltd., London, UK) based AMD64 3200MHz PC-computers or similar. Graphics were designed with Matlab (The MathWorks, Inc., Massachusetts, USA) and Gnuplot 4.2 (http://www.gnuplot.info).

# Results

## Non stimulated fish

To investigate how locomotor and electrical behavior change in long periods we acquired ~ 48 h continuous data from freely moving non stimulated fish. All acquisitions started immediately after introducing the fish in the measurement aquarium and were performed in complete darkness. Each experiment started at a different time of the day. Fish were moved back to the maintenance aquarium with natural illumination at the end of the experiments.

A transient behavior, characterized by low average values of IPIs (high frequency of EODs) (Fig. 4A), was observed at the beginning of all recordings. After the transient the behavior reached a stationary state that persisted throughout the experiment: the IPIs distributions oscillated over time between the same two main peaks: a sharper one corresponding to higher values of IPI and another one corresponding to lower values of IPI. Between 24 and 48 hours of recording there was a period of approximately 12 hours with the predominance of the higher IPIs peak (lower frequency), indicating the influence of a circadian rhythm even with the experiment performed in darkness. In the beginning of the transient, the IPIs presented a much flatter distribution with a single mean, just a little higher than the stationary lower IPIs peak. This initial shape of the distribution changed slowly into the shapes of the stationary state over time, along the transient. The typical transient in response to the new environment lasted several hours, corresponding to a long relaxation time $T_R$, defined as the time needed to the IPIs reach the stationary state. For all fish recorded (N = 4) $T_R$ = (10.2 ± 4.8) h. Despite quantitative differences all fish presented a long initial transient and a final stationary state oscillating between two well defined mean IPIs.

From each experiment data set we also inferred the fish movement by computing the sum of standard deviations of the LIA at each electrode (Fig. 4B). Since the LIA is inversely proportional to the fish-electrode distance, if the fish stands still for a certain time, the LIA will be practically the same from pulse to pulse, providing a low value of standard deviation. Conversely, the amplitude will change from pulse to pulse at least in a few electrodes in an active moving fish, producing higher values of standard deviation for those electrodes. By summing the standard deviations over all electrodes we estimated the fish movement, disregarding its position and direction. Our results pointed that the fish movement was maximum during the initial transient and evolved to a stationary state that alternated periods of vigorous activity and absence of movement (Fig. 4B). The duration of the transient in the movement analysis coincided with that of the transient in EOD activity.

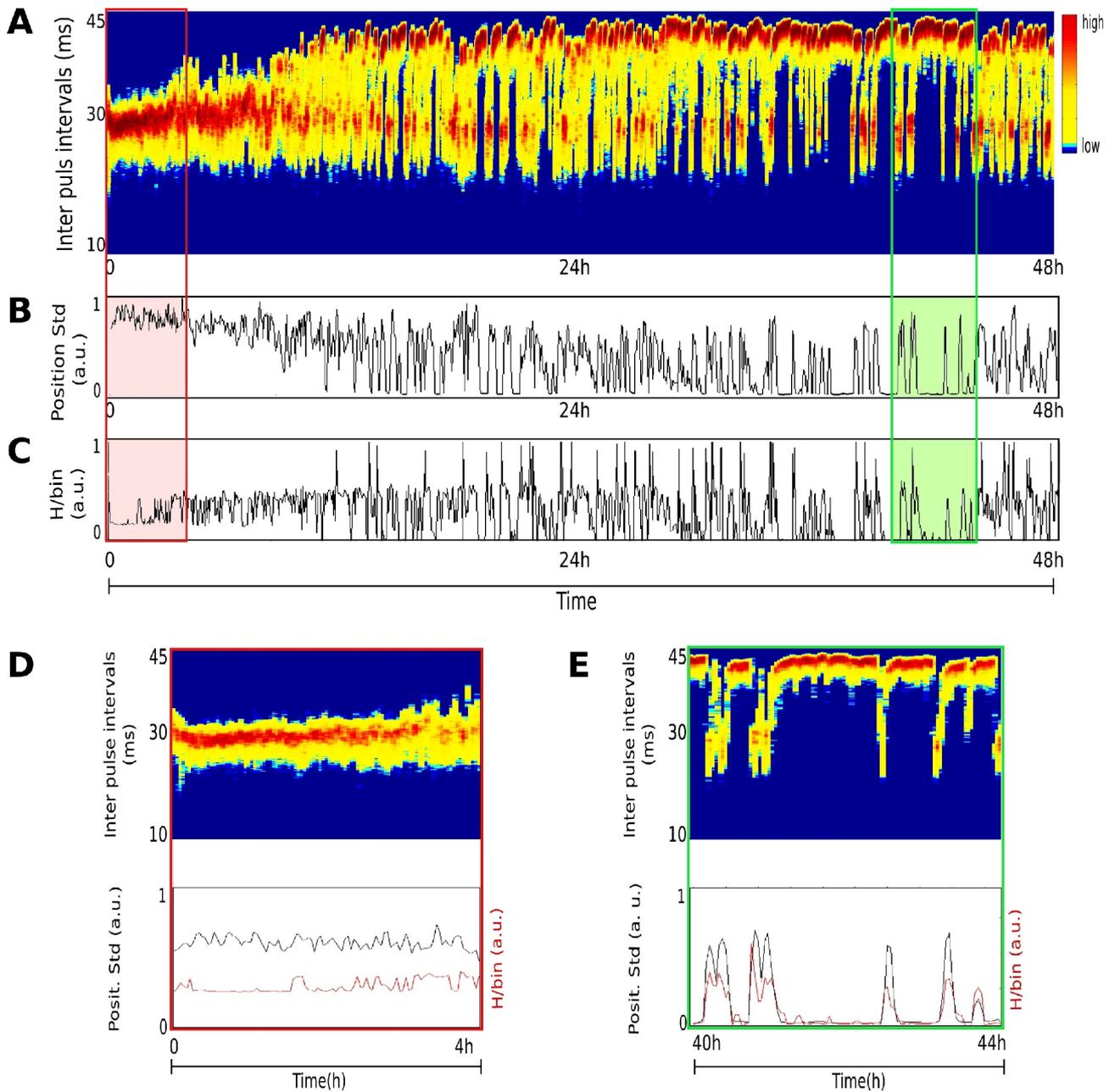

**Figure 4. Non estimulated fish (48 h experiment). A-** Sliding window histogram of IPIs *versus* time. The histograms are calculated in 40s windows. High probabilities are shown in red and low probabilities in blue. There is a transient behavior, at the beginning of the recording, characterized by low average values of IPIs. After 10h the behavior reaches a stationary state that persists throughout the experiment: the IPIs distributions oscillate over time between the same two main peaks of IPIs. **B -** Inferred movement *versus* time. The fish movement was maximum during the initial transient and evolved to a stationary state that alternated periods of vigorous activity and total absence of movement. **C -** Entropy *versus* time. During the transient period the entropy presents intermediate values. After the transient, the entropy oscillates between higher and almost zero values. **D –** Detail during the transient behavior. Above: Sliding window histogram of IPIs *versus* time. Bellow: Inferred movement (black) and entropy (red) *versus* time. At the beginning of the transient there are low IPIs values (high EOD frequencies) with constant values of entropy and restless movement. **E-** Same as D after habituation to the aquarium. In the stationary state, the IPIs oscillate between two mean values. Movement and entropy are entrained.

Analysis of EOD patterns entropy along time (Fig. 4C) revealed the same qualitative behaviors described above. After a long transient period, the entropy followed qualitatively the peaks and valleys of the locomotor activity: peaks (valleys) of locomotor activity correspond to high (low) variability of EOD timing.

Comparing the locomotor activity and the entropy along time at the beginning of the transient (where no absence of movement was found) with those corresponding to the stationary state (were long periods of absence of movement take place) (Figs. 4D and 4E) we found that the same qualitative information can be obtained from both signals.

During the transient we found higher and mostly constant values of locomotor activity and entropy. Along the stationary state both locomotor activity and entropy entrained their peaks and valleys to clearly express changes in the EOD rhythm. When the IPIs distribution along time was dominated by the higher IPI mean, both locomotor activity and entropy vanish. Nevertheless, their peaks reached the same maximum values found during the transient state when the distribution was dominated by the lower IPI mean.

A more detailed movement analysis, based on the signal of the major component of the EOD pulse at each electrode, revealed a preference of staying at the bottom of the aquarium, near the walls, when both locomotor activity and entropy were low (data not shown). No orientation preferences were noted. Similar qualitative behavior was found for all fish.

**Experiments with stimulation**

Both locomotor and electrical behavior changed when we suddenly turned on the stimuli: an EOD timestamp sequence pre-recorded from another (non stimulated) fish (Fig. 5 e 6). By visual inspection we found that the start/end of the stimulation arouse almost all fish. Most of them, quickly approached the area near the artificial fish, some tried to bite it or just kept swimming around. Depending on the fish, the electrical response varied from short or sustained increasing of frequency, to the absence of EODs for minutes. The most common reaction was the decreasing of IPI while increasing their variability (Fig. 5, 6A and 6D).

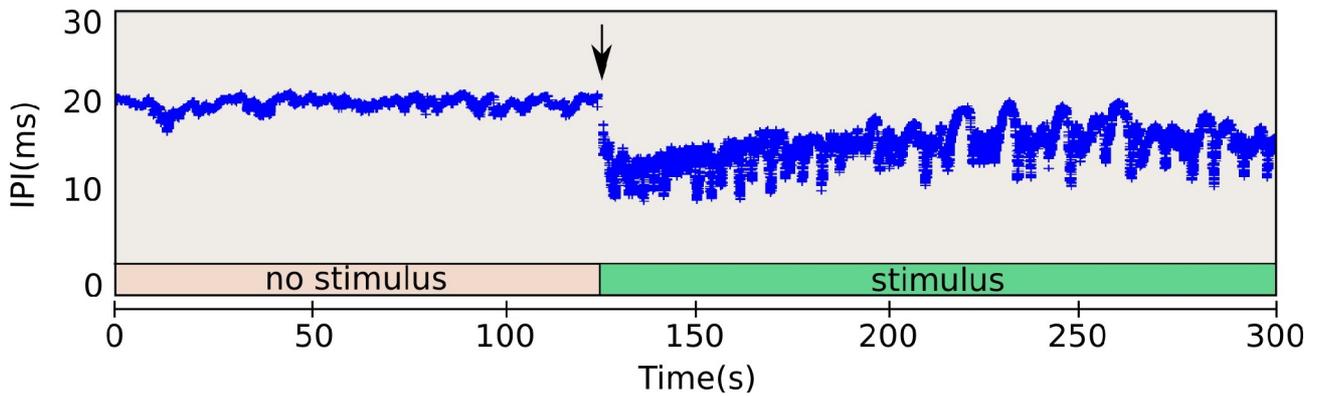

**Figure 5**. **Effect of stimulation in IPIs series.** The arrow shows the moment that the stimuli were turned on. Fish immediately responds to the stimuli by increasing both EOD rate and variability.

To compare the effect of the distinct distributions of stimuli IPIs, a real one (pre-recorded from a fish without stimulation) and a random (flat) distribution of IPIs between 15 and 20 ms, we used two protocols, each one consisting on a sequence of 5 sections of 30 min each. The sequence of sections in one protocol was: control recording (without stimulation) started immediately after introducing a fish in the measurement aquarium, stimulation with a real distribution of IPIs, control, stimulation with random distribution of IPIs, and a final control (Fig. 6); while in the second protocol we inverted the order of presentation of the real and the random stimuli (not shown).

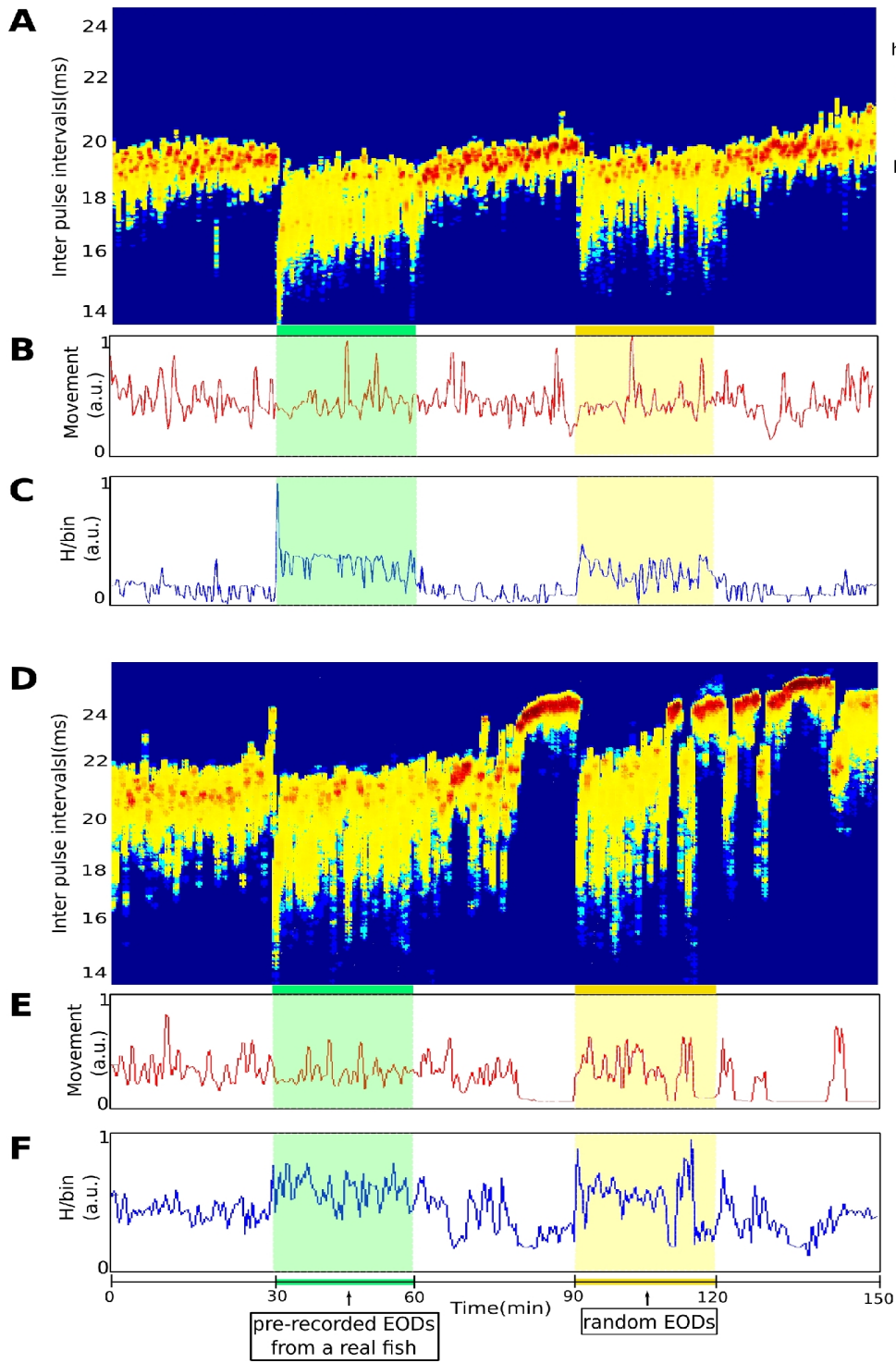

**Figure 6. Experiments with stimulation.** The protocol was a sequence of 5 sections (30 min each) as follows: control recording (without stimulation), stimulation with a real distribution of IPIs (green bar), control, stimulation with random distribution of IPIs (yellow bar), and a final control. In the analysis windows of 40s were considered. **A-** Sliding window histogram of IPIs *versus* time. High (low) probabilities are shown in red (blue). **B -** Inferred movement, and **C -** entropy *versus* time. Fish reacted to both stimuli by decreasing the IPIs values (increasing the EODs frequency) and increasing their variability. During random stimuli (yellow bar) fish presented a slight relaxation with the peak of IPIs tending to higher values over time. The fish was restless moving during the whole experiment with no simple relation to the stimulation sections. The entropy clearly increased (decreased) when both stimuli were turned on (off). **D-F -** same as **A-C** but for an experiment performed 2 weeks later with the same individual. Same qualitative behavior was found with the remarkable exception that during the stimulation with a random distribution fish presented several epochs characterized by: very high IPI values, absence of movement, and very low entropy values, characteristics of behavioral sleep that was expected only during control sections.

During stimulation using the first protocol all fish (N=6) presented a persistent increase in the amplitude of IPI oscillations (typical behavior shown in Fig. 5). All fish showed more discharge interruptions when submitted to the real than to the random stimuli. Spontaneous brief interruptions (IPI > 0,1 s) also happened to non stimulated fish. Five animals out of six (83%) responded to the beginning of all stimuli sections with a transient increase in the EOD frequency. In 3 animals (50%) the EOD frequency increasing for the real stimuli section was two fold that presented for the random stimuli section. In a single case (17%) the real stimuli section evoked dozens of discharge interruptions (~0.5 s each one), but the fish just increased the EOD frequency for the random stimuli section. Qualitatively similar behavior was obtained for fish (N=3) submitted to the second (inverted) protocol.

Fish typically reacted differently when submitted to the stimulation protocol for the first time, just a few days after arriving at the laboratory (Fig. 6 A-C), and after ~ 2 weeks (Fig. 6 D-F). Besides the usual increasing in the EOD frequency during all the real stimuli sections (green bar in Fig. 6 A,D) fish presented, two weeks after arriving at the lab, a distinct reaction to the random stimuli sections (yellow bar in Fig. 6 A, D) by decreasing both EOD frequency and variability. Similar tendency was found in the locomotor behavior (Fig. 6 B, E), as well as in the entropy (Fig. 6 C, F).

Particularly interesting, two weeks after arriving at the lab, both entropy and movement vanished for periods of up to 10 min when the EOD frequency decreased at the end of the random stimuli section (yellow bar in Fig. 6 B,C,E,F). In general, the oscillations on locomotor activity were not directly related to the stimuli sections. Nevertheless, the entropy was clearly higher when the fish was stimulated.

To address whether or not fish manifested clear distinctive behavior when submitted to real or random IPI distributions we built normalized Post Stimulus Time Histograms (PSTH), where we plotted the experimental probability of a fish EOD occurs immediately after an artificial EOD (set as time reference), both for real and random IPI stimuli distributions (Fig. 7). After an artificial EOD from a random stimuli distribution the fish presented a flat EOD probability up to the limit of the average stimulus IPI (~14 ms, Fig. 7). However, after an artificial EOD from a real stimuli distribution, the fish EOD probability smoothly dropped in the interval [1, 5] ms and increased in the interval [6, 12] ms. PSTHs for experiment sections with the real stimuli distribution revealed a fish preference of firing 3% to 5% of the total number of EODs later than observed in PSTHs for experiment sections using the random stimuli distribution.

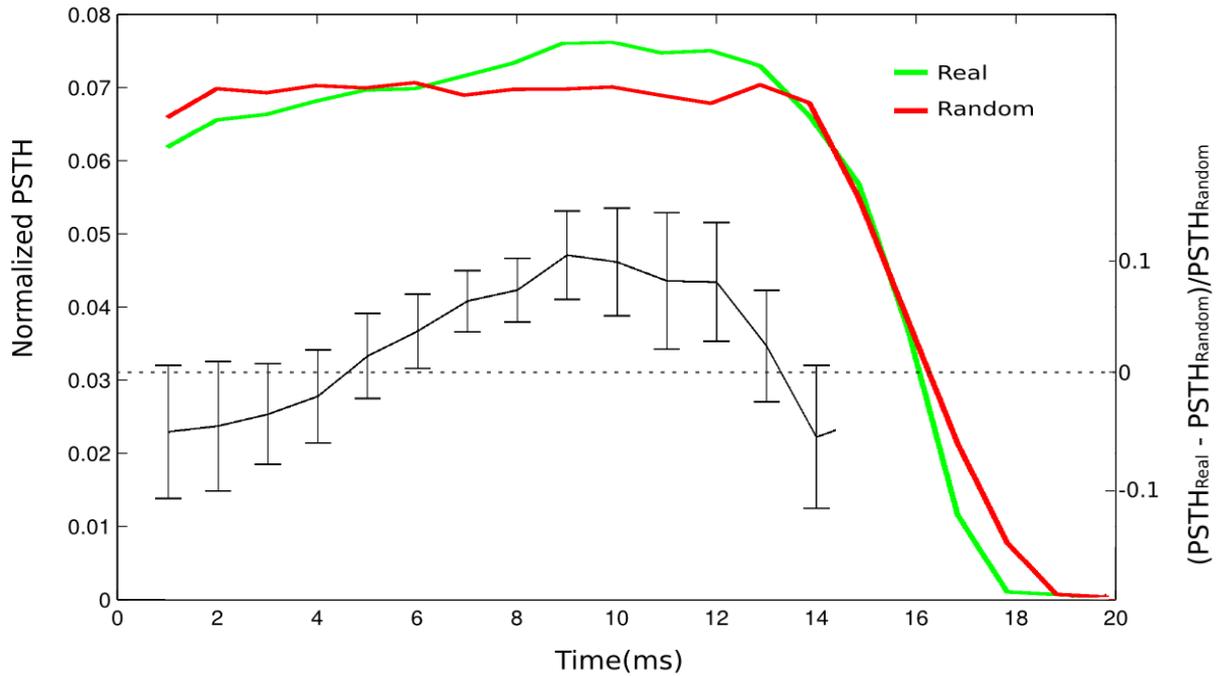

**Figure 7**. **Normalized Post Stimulus Time Histograms (PSTH)**. The PSTH for stimuli from a random distribution of IPIs (red) presents a flat shape up to the limit of the mean IPI (~17 ms), while the PSTH for a real distribution of stimuli IPIs presents a peak of probability around 10 ms. The PSTH change (black) when the random stimuli distribution is replaced by a real one shows an average decrease of probability (~ 5%) in the range [0, 5] ms while the probability in the range [6, 12] ms increases (~ 10%). Error bars are the standard deviations obtained from experiments with 8 fish. The change correspond to 3% to 5% of the total number of EODs and is an evidence that fish discriminate between real and random stimuli distributions (ANOVA, $p<0.002$).

In order to generalize the analysis we calculated both PSTHs for 8 different fish, integrated them in steps of 1 ms and computed how much the PSTH obtained in sections with the real stimuli distribution deviates from the flat PSTH for sections with the random stimuli distribution, i.e., we calculated (PSTHReal - PSTHRandom)/PSTHRandom for each 1 ms bin from 1 to 14 ms (Fig.7). From the average behavior of 8 fish it was clear that in response to an artificial EOD from a real stimuli distribution fish reduced their EOD probability from 1 to 5 ms and increased it from 6 to 12 ms (ANOVA, $p<0.002$).

# Discussion

Pulse-type weakly electric fish allow noninvasive access to central nervous system patterns involved in interesting and not yet completely known tasks: electrolocation and electrocommunication (Lissman,1958; Heiligenberg, 1991; Bullock, 1999).

However, due to the experimental difficulties of dealing with a complex spatio-temporal electrical pattern self-generated by a freely moving fish, most work dealt with interaction between fish with very different pulse amplitudes (Westby, 1975; Arnegard and Carlson, 2005), remarkable progress has been made by restricting the fish movement (Bell et al., 1974) to obtain a simpler temporal pattern of stereotyped electric pulses, which is a completely valid simplification for electrolocation study purposes (Caputi et al., 2002; Bullock et al. 2005; Hitschfeld et al. 2009).

From these important seminal experiments we have learned many details of the pulse generation by the electric organ of the fish (Bullock et al. 2005), as well as how the detection of self- and conspecific- generated signals occurs in the sensitive electroreceptor organs in the skin and is path into the fish brain (Aguilera et al., 2001; Caputi, 2004).

It was also reported that fish usually respond to environmental changes or stimulus (light, electric, mechanic, vibration) with a transient increasing of electric organ discharge frequency (Kramer, 1994), called novelty response (Post and von der Emde, 1999; Aguilera and Caputi, 2002). This poses an important question: if the fish produces an electrical pattern that is sensitive to very subtle perturbations it might also behave in a completely unnatural way while its movements are restricted, precluding any study of its social interaction through electrocommunication. Moreover, the stress caused by manipulating a fish, restricting its movements, or simply moving it to a measurement aquarium could alter the patterns of electrical activity for unknown periods of time.

Recently many advances have been made with new and improved acquisition hardware that allowed researchers to study fishes freely moving in two dimensional (shallow) water tanks (Perrone et al., 2009). The response of the fish was also addressed in studies of social interaction between specimens of a specie that present sexually dimorphic pulse shapes (Wong and Hopkins, 2007; Perrone et al., 2009), due to the difficulties in splitting the pulses of specimens that present very similar pulses.

Here we described the development of an apparatus to address several of these issues. By using a cubic arrangement of electrodes and dedicated real-time software we were able to detect and record the EOD timestamps of a fish while freely swimming in three dimensions during several days, and in the absence of external acoustic, electric, light and mechanical noisy stimuli to avoid undesirable novelty responses. Pulse to pulse variations of the electrical field induced in the electrodes were also recorded and associated to the fish movement. By calculating the entropy of the fish EOD pattern we could compare the electrical and the locomotor behavior. A real-time software allowed us to stimulate the fish in a controlled way, mimicking the pulses of a conspecific but according to a distribution of IPIs chosen by the experimenter among one pre-recorded from an unstimulated real fish, and one with a random (flat) distribution. Besides the electrical shielding that is usual in electrophysiologycal experiments, an efficient acoustic and mechanical insulation was invaluable to allow good recordings: the sensory system of the fish proved to be so sensitive that even small perturbations, such as those provided by closing a door in another floor of the building or someone walking around, were enough to elicit novelty responses.

Our 48h experiments with non-stimulated fish revealed that introducing a fish into a novel environment (the acquisition aquarium) triggered a transient exploratory behavior characterized by high entropy and restless movement all around the aquarium. After a long period (the order of 10 hours) a final stationary state with the EODs oscillating between two mean frequencies was reached. In spite of the experiments been performed in complete darkness we could still see a circadian modulation in the predominance of such frequencies.

It was not expected that locomotor and electrical behavior (entropy of IPIs patterns) could be related since electrogenesis and body movements are controlled by different neural systems: electromotor and skeletomotor command systems (Caputi, 2004). Surprisingly, after the transient period, movement and entropy were entrained. When the fish is moving, faster updates of the electrical images are needed, thus the EOD frequency is increased. During the transient period, in which the fish is stressed and presenting a restless exploratory behavior in the darkness, they need to update the electrical images as fast as possible, so they fire EODs close to the maximum possible frequency in an almost periodic fashion. This behavior results in high values of inferred movement and low values of entropy.

After the transient, when the fish habituate to the environment, they alternate between periods of vigorous but discontinuous movement resulting in high entropy (NRs elicited by the own random movement) and periods with lack of movement (very low entropy values). These periods with the absence of movement and low values of entropy remarkably concentrate in the day-light part of the circadian rhythm, so we inferred they are indicators of behavioral sleep. This way, entropy and movement obtained through this non-invasive method could be used to characterize fish sleep in behavioral experiments.

During stimulation experiments turning on/off the stimulus arouse almost all fish. Most animals attacked the 'artificial fish', showing the same aggressive behavior presented when two fish are put together in the same aquarium without hiding places. Usually fish with aggressive behavior also presented NR while most fish that avoided approaching the stimulus electrode also presented absence of EODs for periods up to minutes in some cases, according to the dominance/submission rules found in natural environment between real animals. These are evidences that fish actually react as if there was a real fish inside the PVC pipe that houses the stimulation electrode.

Both stimuli with real and random distributions of IPIs triggered increases in the EOD frequency, however we found that the increase due to the beginning of a real stimuli section is bigger than the increase due to the random stimuli beginning. During the stimulation with the random IPIs distribution the peak of frequency slowly decays as it happens during the transient period observed in the recordings with non stimulated fish. In experiments with stimulation the entropy and the movement were not correlated, except in a few periods of behavioral sleep, in the second time the fish was exposed to the same protocol.

We also found other differences between the electrical behavior recorded at the first time the fish was exposed to stimulation, just a few days after arriving at the laboratory, and a second experiment repeated after two weeks. In the second one, the same fish usually presented more episodes of behavioral sleep (absence of movement and entropy vanished) during the control non stimulated sections and sometimes even during the random stimuli section, but never during real stimulation. The same qualitative behavior was found by using a protocol where the random stimuli was presented before the real one, indicating the differences observed are due to the nature of the artificial fish IPIs distributions and not due to a stronger response to the stimuli first presented.

Under stimulation, fish perfomed both electrolocation and electrocommunication, therefore entropy and movement were mostly not correlated. Moreover, some mechanism of JAR could also be present. To clarify this issue we compared several fish EOD occurence probabilities immediately after being exposed to a real or a random stimulus in PSTHs. The probability of an EOD after a pulse coming from a random distribution of IPIs is constant over time. However, the probability of an EOD after a stimulus pulse coming from a real distribution of IPIs is slightly reduced (3-5% of the pulses) for short times and increased for longer times. This result is a clear evidence that fish is able to discriminate between these two distributions. Moreover, the result obtained with the real stimuli agrees with those preliminary found in an experiment with two real fish hidden in plastic pipes (Westby, 1979). We also built PSTHs for both real and random IPIs distributions using finer time bins (not shown) to analyze the probability around time=0 (coincidences), and there were no evidences of a JAR mechanism.

It has been already hypothesized that electric fish might separate their own electric pulses from external ones (Szabo et al, 1975; Caputi and Nogueira, 2012). Recently it was shown, in brain slices, that spherical neurons at the fast electrosensory path of *Gymnotus omarorum* have an intrinsic refractory time of ~10 ms which is about half of the average IPIs in those fish (Nogueira and Caputi, 2011). In the same work there are also evidences, from chronically implanted fish experiments, that the animal uses this refractory time to minimize the interference of conspecific pulses in the sensory pathway. Our PSTH results revealed a preference for discharging after the half of the average IPI of the conspecific (stimuli from real distribution), which means after the conspecific refractory period. In other words, our results point that in our experimental conditions the fish changes the timing of the EOD to be detected by the conspecific that would have to delay its next EOD to avoid the refractory period of its own spherical neurons.

We propose that a fish estimates the refractory period of a conspecific and choose to discharge within or out of this period. For example, to electrolocate without been detected it would be interesting to discharge within the refractory period; whereas to communicate a fish would discharge out of the conspecific refractory period. If this is the case, one must take in account both refractory periods when analyzing communication between fish.

To confirm this hypotheses and to estimate the refractory period in freely moving fish Gymnotus carapo, we are currently planning real-time experiments in which stimulus pulses will be delivered according to the fish's own EODs, as well as experiments with two fish in the same tank and software techniques to classify and split their pulses.